# Visible emission spectroscopy of highly charged tungsten ions in LHD I: Survey of New Visible Emission Lines

M. Shinohara[1], K. Fujii[1*], D. Kato[2,3], N. Nakamura[4],
M. Goto[2,3], S. Morita[2,3], M. Hasuo[1], LHD Experiment Group[2]
[1]Department of Mechanical Engineering and Science, Graduate School of Engineering, Kyoto University, Kyoto 615-8540, Japan
[2]National Institute for Fusion Science, Toki 509-5292, Japan
[3]Dept. of Fusion Science, The Graduate University for Advanced Studies, Toki 509-5292, Japan
[4]Institute for Laser Science, The University of Electro-Communications, Tokyo 182-8585, Japan
fujii@me.kyoto-u.ac.jp

**Abstract**

We found 12 unknown visible emission lines from the core plasma of Large Helical Device with highly charged tungsten ions accumulated. The observation was made with our home-built échelle spectrometer, which covers the wavelength range of 450–715 nm with a wavelength resolution of < 0.05 nm for two lines of sight; one line passes both the core and edge plasmas and the other passes only the edge plasma. These emission lines are attributed to highly charged tungsten ions because (1) they were observed only after a tungsten pellet was injected into the plasma, (2)they were observed only from the core plasma where the electron temperature is 1 keV, (3) they show line broadenings that are close to the Doppler widths of tungsten ions with 1 keV temperature and (4) the wavelengths of some of these emission lines are close to the calculation results for tungsten ions in the charge state of 25–28.

## 1. Introduction

Tungsten (W) is planned for use as the first surface material of the divertors in ITER because of its refractory properties, low hydrogen retention, and high thermal conductivity [1,2,3,4]. Since tungsten is a high-*Z* element (*Z* = 74), where *Z* is atomic number, it still has bounded electrons in the ITER core plasma. With an electron temperature of 10–20 keV, highly charged tungsten ions in the ITER core plasma radiate intense extreme ultraviolet (EUV) and X-ray radiation, causing radiation loss.

Spectroscopic observations of tungsten transport in the plasma has been started for tokamaks equipped with tungsten limiters [5,6], and have recently been expanded to large-scale tokamaks and stellarlators [7,8,9,10,11,12]. Since the resolving power, defined as the ratio of the observed wavelength to the instrumental width of the spectrometer, is typically $10^2$ in the EUV and soft X-ray regions, many emission lines from the different charge states are blended. In order to evaluate the contribution of each charge state, the observed spectra have been compared with synthetic spectra [8-10] constructed from calculated emission-line wavelengths and transition probabilities, as well as the population of the upper states, using atomic data models such as Flexible Atomic Code [13]. Therefore the accuracy of these models is essential and efforts to sophisticate them have been made [14,15].

Some highly charged tungsten ion emission lines resulting from electric-dipole-forbidden transitions are expected to appear in the visible range [16]. One advantage of visible spectroscopy is that a resolving power of $10^4$ is easily achieved. Another advantage is that the use of glass optics, including windows, lenses, and optical fibers, for emission transfer, which enables a flexible instrumental layout. However, only a few visible emission lines have been observed or identified for highly charged tungsten ions.

Studies aimed at discovering and identifying visible and near ultraviolet (UV) emission lines have been performed by electron beam ion trap (EBIT) experiments [17,18,19,20,21,22] and theoretical calculations [23,24,25]. In the EBIT experiment, tungsten ions in a desired charge state can be generated by a mono-energetic electron beam. Table 1 shows a list of the emission lines of tungsten ion with the charge state, $q$ = 13–28 in the wavelength $\lambda$ = 450–715 nm experimentally observed for EBIT experiments. The theoretical calculation results are also shown in this table. Although we list most of the calculated results in this wavelength range, they are limited to $q$ = 13, 25, 26, 27, and 28; theoretical calculations for the other charge states are still unavailable.

Table 1 Theoretical and experimental wavelengths and charge states of highly charged tungsten ions in $\lambda$ = 450–715 nm. The corresponding transitions of $W^{13+}$ and $W^{25+}$ - $W^{28+}$ are listed in the calculation column. “-” and “n.d.” indicate that no observations were made and that observations were made but the emission was *not detected* with sufficient signal to noise ratio,” respectively. The identified transitions are indicated by the arrows in the 4th column.

| Calculation | | | Experiment in EBIT | | Experiment in the LHD |
|---|---|---|---|---|---|
| *q* | Upper- Lower levels | Wavelengths (nm) | *q* | Wavelengths (nm) | Wavelengths (nm) |
| | | | 15 | 450.23 [19] | n.d. |
| | | | 21 | 450.70 [19] | n.d. |
| | | | 25 | 451.15 [19] | n.d. |
| | | | 21 | 451.17 [19] | n.d. |
| | | | 16 | 455.52 [19] | n.d. |
| | | | 19 | 456.43 [19] | n.d. |
| | | | 13 | 457.26 [19] | n.d. |
| | | | 13 | 459.08 [19] | n.d. † |
| | | | 23 | 459.25 [19] | n.d. |
| | | | 21 | 459.99 [19] | n.d. |
| | | | 20 | 462.40 [19] | n.d. |
| | | | 14 | 462.59* [19] | 462.64 |
| 13 | $[[(4f^5_{5/2})_{5/2}(4f^7_{7/2})_{7/2}]_5 5p_{1/2}]_{9/2}$ $\rightarrow[(4f^6_{7/2})_6 5p_{1/2}]_{11/2}$ | 460.93[22] | ←13 | 462.64* [22] | |
| | | | 21 | 463.50 [19] | n.d. |
| 26 | $[4d^{10}4f_{7/2}{}^2]_6$ $\rightarrow[4d^{10}4f_{5/2}4f_{7/2}]_5$ | 467.79 [23]<br>464.7 [20] | ←26 | 464.63** [19]<br>464.636 [20] | n.d. |
| | | | 25 | 467.59 [19] | n.d. † |
| | | | 24 | 467.80 [19] | n.d. |
| | | | 24 | 468.22 [19] | n.d. † |
| | | | 21 | 468.39 [19] | n.d. † |
| | | | 25 | 469.21 [19] | n.d. |
| | | | 24 | 471.18 [19] | n.d. |
| 26 | $[4d^{10}4f_{5/2}4f_{7/2}]_6$ $\rightarrow[4d^{10}4f_{7/2}{}^2]_4$ | 472.16 [23] | | | |
| | | | 16 | 472.39 [19] | n.d. |
| | | | 13 | 472.68 [19] | n.d. |
| | | | 19 | 474.49 [19] | n.d. |
| 26 | $[4d^{10}4f_{7/2}{}^2]_4$ $\rightarrow[4d^{10}4f_{7/2}{}^2]_6$ | 482.66 [23] | | | |
| | | | 13 | 486.33 [22] | n.d. † |
| | | | 25 | 493.62 [19] | 493.6 |
| 25 | $[4d^{10}(4f^2_{5/2})_4 4f_{7/2}]_{11/2}$ $\rightarrow[4d^{10}4f^3_{7/2}]_{9/2}$ | 494.66 [21] | ←25 | 493.84 ± 0.15 [20] | n.d. |
| 27 | $[4d_{5/2}{}^{-1}4f_{7/2}{}^2]_{11/2}$ $\rightarrow[(4d_{5/2}{}^{-1}4f_{5/2})_1 4f_{7/2}]_{13/2}$ | 498.4 [24] | | n.d. | 498.92 |
| | | | | n.d. | 499.90 |
| 26 | $[4d^{10}4f_{5/2}4f_{7/2}]_3\rightarrow[4d^{10}4f_{5/2}{}^2]_2$ | 501.80 [23]<br>502.3 [20] | ←26<br>←26 | 501.99 ***[19]<br>502.153 ***[20] | 501.99 |
| 26 | $[4d^{10}4f_{5/2}4f_{7/2}]_4\rightarrow[4d^{10}4f_{5/2}{}^2]_2$ | 509.09 [23] | | -<br>- | 509.11<br>509.81 |
| 26 | $[4d^{10}4f_{7/2}^{2}]_2 \rightarrow [4d^{10}4f_{5/2}4f_{7/2}]_1$ | 516.01 [23] | | | |
| | | | 13 | 517.77 [22] | n.d. |
| 13 | $[[(4f^5_{5/2})_{5/2}(4f^7_{7/2})_{7/2}]_3 5p_{1/2}]_{3/2} \rightarrow [[(4f^5_{5/2})_{5/2}(4f^7_{7/2})_{7/2}]_3 5p_{1/2}]_{5/2}$ | 527.11 [22] | ←13 | 527.74 [22] | n.d. |
| 26 | $[4d^{10}4f_{5/2}^2]_2 \rightarrow [4d^{10}4f_{5/2}^2]_4$ | 536.67 [23] | | -<br>- | 537.61<br>539.63 |
| 13 | $[[(4f^5_{5/2})_{5/2}(4f^7_{7/2})_{7/2}]_4 5p_{1/2}]_{9/2} \rightarrow [(4f^6_{7/2})_4 5p_{1/2}]_{9/2}$ | 543.79 [22] | ←13 | 546.23 [22] | n.d. |
| 13 | $[(4f^5_{5/2})_{5/2}5s^2]_{5/2} \rightarrow [(4f^5_{5/2})_{7/2}5s^2]_{7/2}$ | 552.08 [22] | ←13 | 549.95 [22] | n.d. † |
| | | | | - | 579.75 |
| 27 | $[(4d_{5/2}^{-1}4f_{5/2})_5 4f_{7/2}]_{11/2} \rightarrow [(4d_{5/2}^{-1}4f_{5/2})_3 4f_{7/2}]_{13/2}$ | 582.8 [24] | | | |
| 27 | $[4d_{5/2}^{-1}4f_{7/2}^2]_{11/2} \rightarrow [4d_{5/2}^{-1}4f_{5/2}^2]_{13/2}$ | 583.3 [24] | | | |
| 27 | $[4d_{5/2}^{-1}4f_{7/2}^2]_{15/2} \rightarrow [(4d_{5/2}^{-1}4f_{5/2})_5 4f_{7/2}]_{13/2}$ | 584.1 [24] | | - | 585.56 |
| 27 | $[(4d_{5/2}^{-1}4f_{5/2})_3 4f_{7/2}]_{11/2} \rightarrow [(4d_{5/2}^{-1}4f_{5/2})_5 4f_{7/2}]_{15/2}$ | 587.6 [24] | | | |
| 25 | $[4d^{10}\ 4f_{5/2}(4f^2_{7/2})_6]_{13/2} \rightarrow [4d^{10}\ (4f^2_{5/2})_4 4f_{7/2}]_{11/2}$ | 588.13 [21] | ←25 | 587.63 [22] | n.d. † |
| 27 | $[(4d_{5/2}^{-1}4f_{5/2})_2 4f_{7/2}]_{11/2} \rightarrow [4d_{5/2}^{-1}4f_{5/2}^2]_{9/2}$ | 600.7 [24] | | | |
| 28 | $[4f_{5/2}^9, 4f_{5/2}]_5 \rightarrow [4f_{5/2}, 4f_{7/2}]_6$ | 605.57 [21] | | | |
| | | | | - | 620.27 |
| 27 | $[4d_{5/2}^{-1}4f_{5/2}^2]_{13/2} \rightarrow [(4d_{5/2}^{-1}4f_{5/2})_4 4f_{7/2}]_{13/2}$ | 669.3 [24] | | n.d. [26] | 668.89 |
| 26 | $[4d^{10}4f_{5/2}4f_{7/2}]_1 \rightarrow [4d^{10}4f_{5/2}4f_{7/2}]_2$ | 685.16 [23] | | | |
| 27 | $[(4d_{5/2}^{-1}4f_{5/2})_4 4f_{7/2}]_{13/2} \rightarrow [(4d_{5/2}^{-1}4f_{5/2})_4 4f_{7/2}]_{11/2}$ | 701.4 [24] | | | |

* Slightly different wavelengths and different charge states were reported for this emission line by two groups; Komatsu et al claimed $\lambda_0$ = 462.59±0.05 nm and $q$ = 14 [19] while Zao et al claimed $\lambda_0$ = 462.64 nm and $q$ = 13 [20].

** The wavelength of this emission line had been incorrectly reported as 464.41 nm in [23] and was corrected in [19].

*** Different values were reported by the two groups for the wavelength of this emission line.

† This emission line was not detected due to the overlap by other intense emission lines.

The characteristic of plasma in an EBIT experiment significantly contrasts that in a magnetic fusion device. In a magnetic fusion device, thermal hydrogen/deuterium plasma is generated; the electrons have a Maxwell energy distribution and high temperature protons/deutrons exist. In the fusion plasmas, electron density is $10^{19}$–$10^{21}$ $m^{-3}$, whereas that in the EBIT is on the order of $10^{16}$–$10^{18}$ $m^{-3}$ [ 27 , 28 ]. Kato et al demonstrated that a near-UV emission line of $W^{26+}$ $[4d^{10}4f_{5/2}4f_{7/2}]_5 \rightarrow [4d^{10}4f_{5/2}{}^2]_4$ (magnetic-dipole transition, central wavelength, $\lambda_0$ = 389.4 nm) is also observed in the hydrogen thermal plasma generated in Large Helical Device (LHD) [29]. They have also measured the spatial distribution of the emission intensity [29]. In the present study, we utilize LHD plasmas to survey the visible emission lines resulting from highly charged tungsten ions found that are intense in hydrogen thermal plasmas.

## 2. Experimental Setup

LHD is a heliotron-type plasma-confinement device. Its toroidal and poloidal cross sections and our definition of *XYZ*-axes are given in Fig. 1(a) and (b), respectively. Hydrogen plasma with a volume of ~30 $m^3$ is confined by a pair of superconducting twisted coils. A normalized minor radius, $\rho$, which is a measure of the closed magnetic flux surface, is indicated in Fig. 1(a). The magnetic field strength at the plasma center ($\rho \sim 0$) is 2.75 T and points nearly parallel to the $Y$ direction.

The first walls and the divertor plates of LHD are made of stainless steel (SUS316L) and carbon composite, respectively; no tungsten is used in the plasma-facing components. The plasma is heated by electron cyclotron wave and neutral beam injection. The radial distributions of the electron temperature, $T_e$, and density, $n_e$, are measured by the Thomson scattering method (spatial resolution: ~3 mm, temporal resolution: 30 ms, accuracy: ~10%) [30].

For the study of the tungsten transport in LHD plasma, pellet injection experiments have been performed [31,32]. The tungsten pellet is made from a tungsten wire, 0.15 mm in diameter, covered by a polyethylene tube, as shown in the inset of Fig. 1(a). The pellet is accelerated by 18 atm helium gas in a 6 m long acceleration tube equipped with a pneumatic pipe-gun system.

Temporal evolutions of $T_e$ and $n_e$ at the plasma center are shown in Fig. 2(a). The tungsten pellet is injected at $t$ = 4.05 s. After the injection, $n_e$ increases and $T_e$ decreases. The radial distributions of $T_e$ and $n_e$ before ($t$ = 3.7 s) and after ($t$ = 4.5 s) the injection are shown in Fig. 2(b). Ion temperature, $T_i$, is not measured in this experiment, but $T_i \approx T_e$ can be assumed in this electron density range. Although the tungsten ion density distribution measurement is unavailable, the $n_e$ profile and bremsstrahlung intensity measurement suggest that the tungsten ions are widely distributed between the edge and core region just after pellet injection, and then gradually diffuse out. Since the typical $T_e$ in LHD core plasma is ~1 keV, mainly $q < 30$ tungsten ions are generated [15]. These ions are heated by Coulomb collision with thermal electrons and protons in the plasma. $T_e$ in the edge region ($\rho > 1$) is less than 300 eV.

Emission from the plasma is observed from two lines of sight (LOSs, arrows in Fig. 1(b)). One LOS (LOS1) observes only the edge of the plasma, whereas the other (LOS2) observes both the edge and core regions. The diameter of each LOS in the plasma is roughly 4 cm. Emissions from the plasma are focused by convex lenses (fused silica, focal length: 30 mm) on the edges of two optical fibers (core diameter: 300 m; length: ~50 m). The fibers are connected to an optical fiber bundle (core diameter: 100 μm) and introduced into an échelle spectrometer, which covers the visible emission in the 450–715 nm wavelength range, with a wavelength resolution of ~0.05 nm. The details of this spectrometer are described in the Appendix.

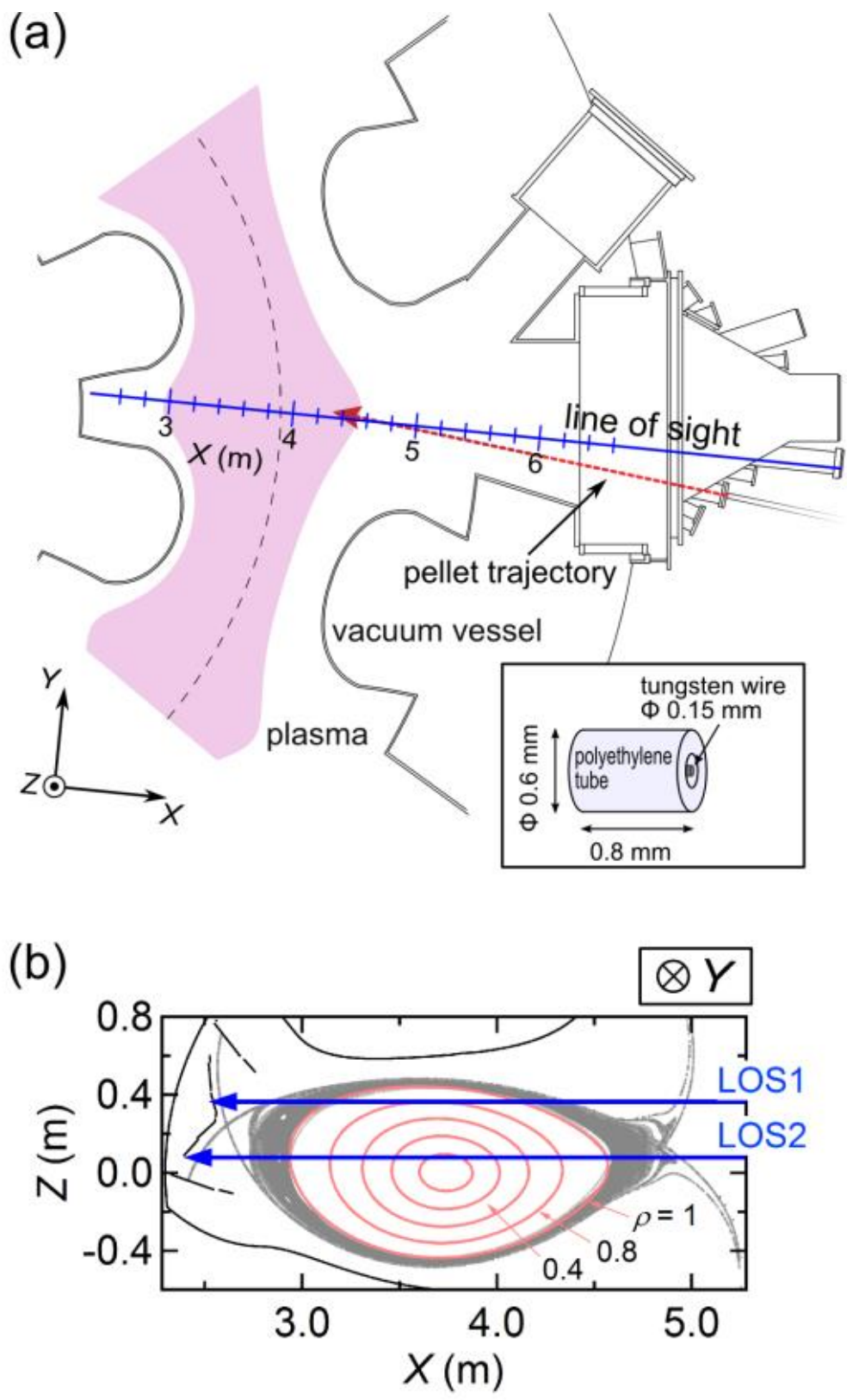

Fig. 1 (a) Toroidal and (b) poloidal cross sections of LHD and lines of sight (blue lines) for the emission observation. The magnetic flux surfaces are plotted by ellipsoidal curves in (b) with the normalized minor radius $\rho$. The inset in (a) is a schematic of the tungsten pellet.

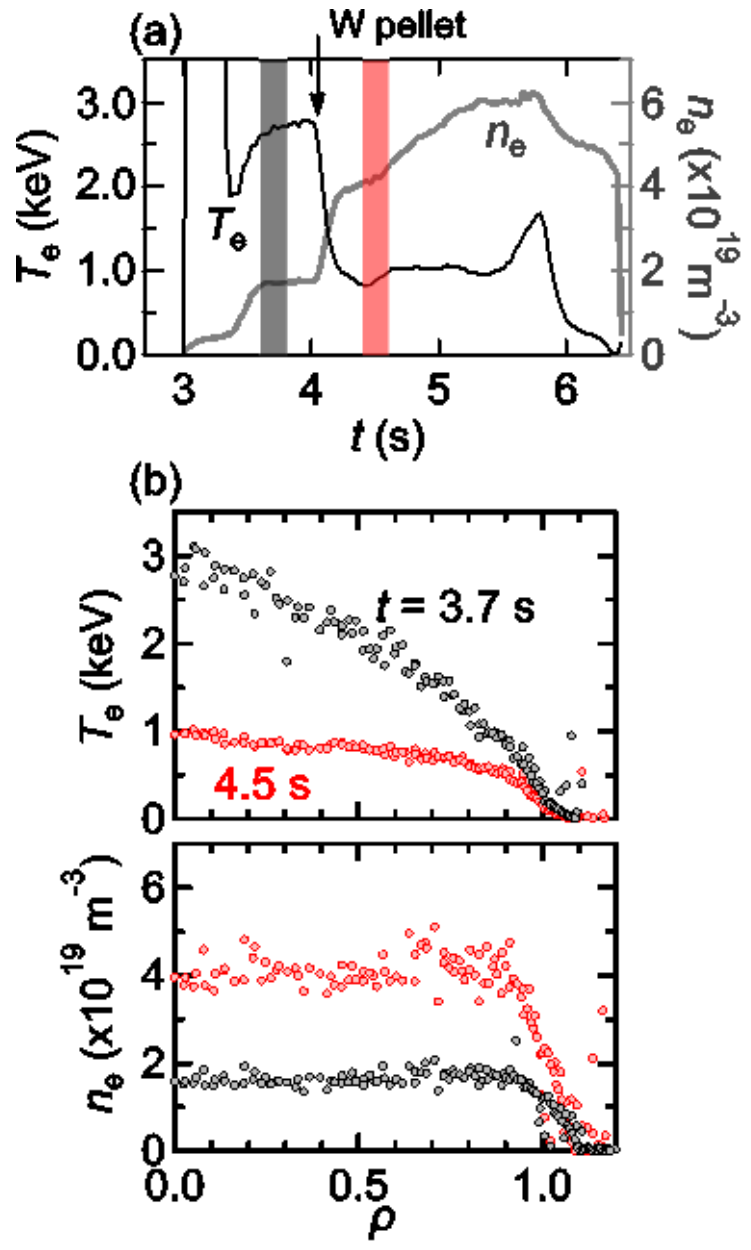

Fig. 2 (a) Temporal evolutions of $T_e$ and $n_e$ at the plasma center. The timing of the pellet injection is shown by a vertical arrow. The gray and red stripes indicate the exposure times of the emission observation before and after pellet injection, respectively. (b) Radial distributions of $T_e$ (upper figure) and $n_e$ (lower figure) measured before ($t$ = 3.7 s) and after ($t$ = 4.5 s) the pellet injection.

## 3. Results

The spectra observed for LOS1 at $t$ = 3.6–3.8 s and 4.4–4.6 s are indicated, on a logarithmic scale, by the gray and red curves in Fig.3, respectively. The exposure time is 200 ms. Intense emission lines of hydrogen atoms, helium atoms and carbon ions are shown in the spectra. Besides these intense lines, a noise-like base line is also seen. It is not a noise but an aggregate of a number of weak emission lines of hydrogen molecules and other impurity ions, including carbon, iron, and chromium, which are difficult to distinguish in this figure (see close-up spectra, Fig.4 and Fig. 5 later). A continuum emission is also observed in the baseline, which is mainly due to the bremsstrahlung. Note that the spectral noise amplitude is less than $10^{-3}$ W $m^{-2}sr^{-1}nm^{-1}$. After pellet injection, the line intensities observed for both LOS1 and LOS2 increase by factors of roughly 1.6 and 1.4, respectively, mainly due to the increase in the electron density. The continuum intensity also increases due to an increase in electron density and the effective ionic charge.

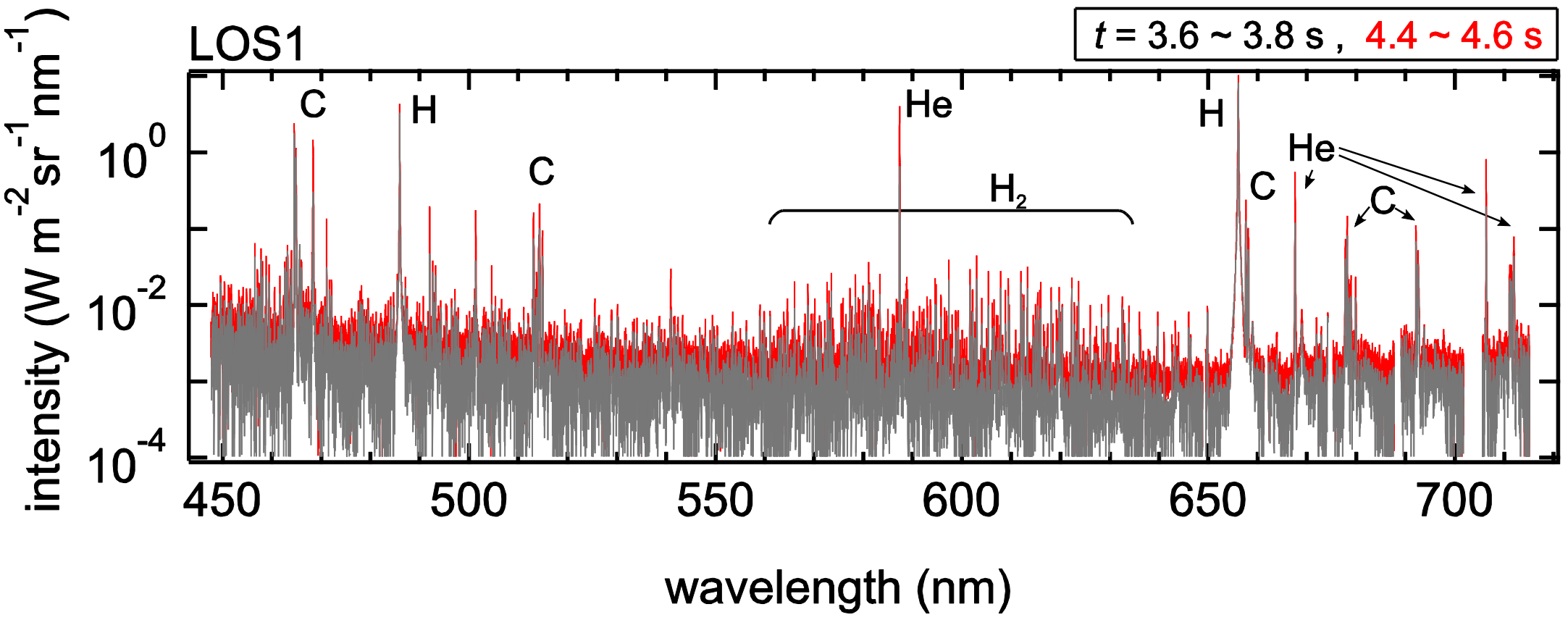

Fig.3 (a) Spectra observed before ($t$ = 3.6–3.8 s) and after ($t$ = 4.4–4.6 s) the pellet injection for LOS1. Some of molecular, atomic, and ionic emission lines of hydrogen, helium and carbon are indicated.

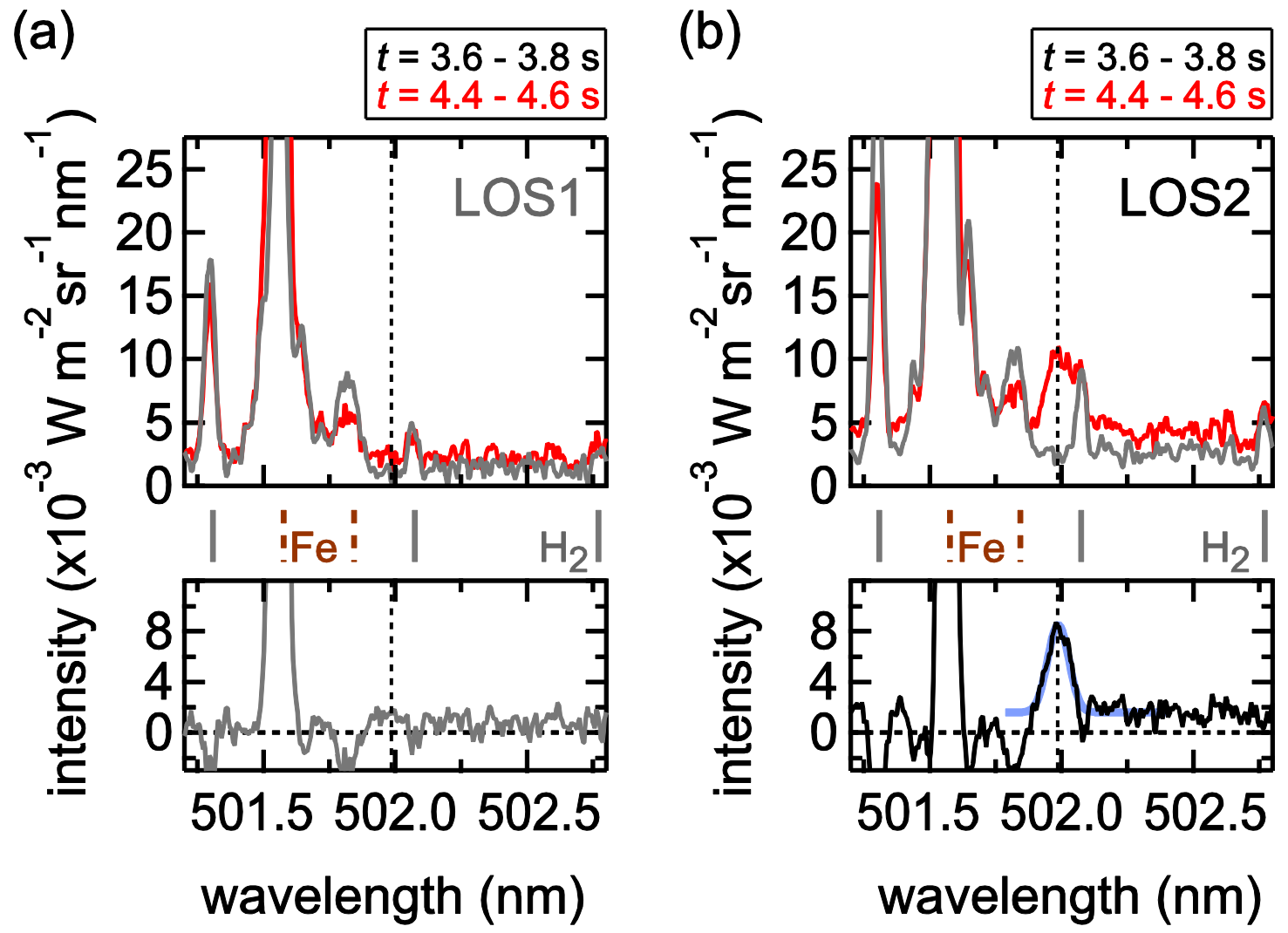

Fig.4 (Upper panels) Close-up spectra observed for (a) LOS1 and (b) LOS2 before ($t$ = 3.6–3.8 s, gray curves) and after ($t$ = 4.4–4.6 s, red curves) the pellet injection. The intensities of the spectra observed at $t$ = 3.8–4.0 s for LOS1 and LOS2 are magnified by a factor of 1.6 and 1.4, respectively. (Lower panels) Differential spectra between the spectra observed at $t$ = 4.4–4.6 s and $t$ = 3.6–3.8 s. The center wavelength of the $W^{26+}$ $[4d^{10}4f_{5/2}4f_{7/2}]_5 \rightarrow [4d^{10}4f_{5/2}^{\,2}]_4$ transition ($\lambda_0$ = 501.99 nm) is indicated by a vertical dotted line. The blue curve in the lower panel in (b) is a fit result of the differential spectrum observed for LOS2 by a Gaussian function. The central wavelengths of other emission lines from hydrogen molecules and iron ions are given by the gray and brown dashed bars, respectively.

The spectra observed for LOS1 and LOS2 in the 501.0–503.0 nm wavelength range are shown in the upper panels of Fig.4(a) and 4(b), respectively. The central wavelengths of emission lines of hydrogen molecules and iron ions are indicated by the vertical bars between the upper and lower panels. The spectra observed at $t$ = 3.6–3.8 s for LOS1 and LOS2 in Fig. 4 are scaled by factors of 1.6 and 1.4, respectively, which roughly compensates for the intensity increase in the hydrogen molecule and impurity lines after pellet injection. The differential spectrum between the spectra observed at $t$ = 4.4–4.6 s and $t$ = 3.6–3.8 s for LOS1 and LOS2 are shown in the lower panels of Fig.4 (a) and (b), respectively. The vertical dotted line shows the central wavelength ($\lambda_0$ = 501.99 nm) of the emission line due to $[4d^{10}4f_{5/2}4f_{7/2}]_3 \rightarrow [4d^{10}4f_{5/2}{}^2]_2$ transition of $W^{26+}$ [19,23]. It is clear that this emission line appears only in the spectrum observed for LOS2 after the pellet injection. This spatial dependence of the emission appearance indicates that highly charged tungsten ions with $q$ ~ 26 locally exist in LHD core plasma. A similar spatial dependence of the tungsten emission intensity in LHD plasma has previously been reported for another transition ($W^{26+}$ $[4d^{10}4f_{5/2}4f_{7/2}]_5 \rightarrow [4d^{10}4f_{5/2}{}^2]_4$, $\lambda_0$ = 389.4 nm) [29].

This 501.99 nm emission line is broader than the other emission lines of hydrogen molecules and iron ions, and the instrumental width (0.032 nm at this wavelength). We fit the differential spectrum by a Gaussian function, as shown by the blue curve in Fig. 4, with adjustable parameters of area, central wavelength, width, and baseline level. The full width at half maximum (FWHM) of the emission line is observed to be 0.093 nm. By subtracting the effect of the instrumental width, the FWHM is estimated to be 0.086 nm. The line broadening may be caused by the Doppler effect, resulting from thermal motion of the tungsten ions, and by the Zeeman split, due to the magnetic field for the plasma confinement. Although the Zeeman split profile and amplitude is unknown for this emission line, we estimate the broadening due to the Zeeman effect as ~0.03 nm by assuming the normal Zeeman effect and $g$ = 0.94 as those for the 668.9 nm line reported in a separate publication [33]. We calculate the tungsten ion temperature to be 0.87 keV by subtracting the Zeeman effect from the observed width. If the Zeeman effect is neglected, the calculated tungsten ion temperature is 0.93 keV. These values are close to $T_e$ in the core region (~1 keV). This suggests that the highly charged tungsten ions are heated close to thermal equilibrium with the electrons and protons in the core region.

As shown in Table 1, some tungsten ion emission with $q$ = 13–26 have been observed in the EBIT experiment in the 450–590 nm wavelength range. In Fig. 5 we show the observed spectra for LOS2 (upper panel) and the differential spectra (gray and black curves in the lower parts, respectively) around the reported wavelengths. The reported wavelengths are indicated by the vertical dotted lines. The central wavelengths of emission lines of hydrogen molecules and impurity ions are also indicated by the vertical bars. We note that the peak at $\lambda$ = 669.2 nm (a square in Fig. 5 (h)) is caused by the image broadening of the intense Balmer-α emission of hydrogen atoms as shown later in Fig.7 (c).

An increase in intensity is observed at $\lambda$ = 462.64 nm, 493.6 nm, and 501.99 nm (vertical dotted lines with a + mark above in Fig. 5). Some emission lines (vertical dotted lines with an asterisk above) are not detected because of the overlap from other intense emission lines. No intensity increases are detected for other wavelength ranges.

As indicated by blue vertical chain lines in Fig. 5 (a)–(h), some emission lines, which have not been observed in the EBIT experiment, are only observed for LOS2 and after the pellet injection. These emission lines have broad profiles similar to the 501.99 nm line. We fit the differential spectra by a Gaussian function and estimate their central wavelengths and line widths. The results are listed in Table 2. The widths including the Doppler and the Zeeman effects assuming the normal Zeeman effect and the $g$ factor same with that of the 668.90 nm line ($g$ = 0.94) [33] are also shown in the table.

The observed widths are close but slightly larger than the calculated widths. The slight difference may be due to the underestimation of their Zeeman profile and/or the $g$ factors. However, at the present stage, it is difficult to evaluate the Zeeman profile or the $g$ factors for these lines from the observed spectra due to the small signal to noise ratio and insufficient wavelength resolution. Further observation with the polarization resolution as well as higher wavelength resolution is necessary.

Table 2. Central wavelengths (1st row) and widths (FWHM, the instrumental widths are already subtracted, 2nd row) of the observed emission lines. The calculated Doppler widths from 1 keV tungsten ions and the calculated widths including the Doppler and Zeeman effects assuming $g$ = 0.94 and 1 keV tungsten temperature are shown in 3rd and 4th rows, respectively. All values are listed in nm.

| Central wavelengths | 462.64 | 493.6 | 498.92 | 499.90 | 501.99 | 509.11 | 509.81 |
|---|---|---|---|---|---|---|---|
| Observed widths | 0.13 | 0.17* | 0.11 | 0.09 | 0.086 | 0.07* | 0.15* |
| Doppler widths for 1 keV $W^{q+}$ | 0.083 | 0.088 | 0.089 | 0.090 | 0.090 | 0.091 | 0.091 |
| Widths including Doppler and Zeeman effect for 1 keV $W^{q+}$ | 0.086 | 0.092 | 0.093 | 0.094 | 0.094 | 0.094 | 0.096 |

| Central wavelengths | 537.61 | 539.63 | 579.75 | 585.56 | 620.27 | 668.89 |
|---|---|---|---|---|---|---|
| Observed widths | 0.16 | 0.15 | 0.17 | 0.15 | 0.16 | 0.16 |
| Doppler widths for 1 keV $W^{q+}$ | 0.096 | 0.097 | 0.104 | 0.105 | 0.111 | 0.120 |
| Widths including Doppler and Zeeman effect for 1 keV $W^{q+}$ | 0.101 | 0.102 | 0.110 | 0.111 | 0.120 | 0.130 |

* The observed widths include ~ 50 % uncertainty due to the unclear spectral profile.

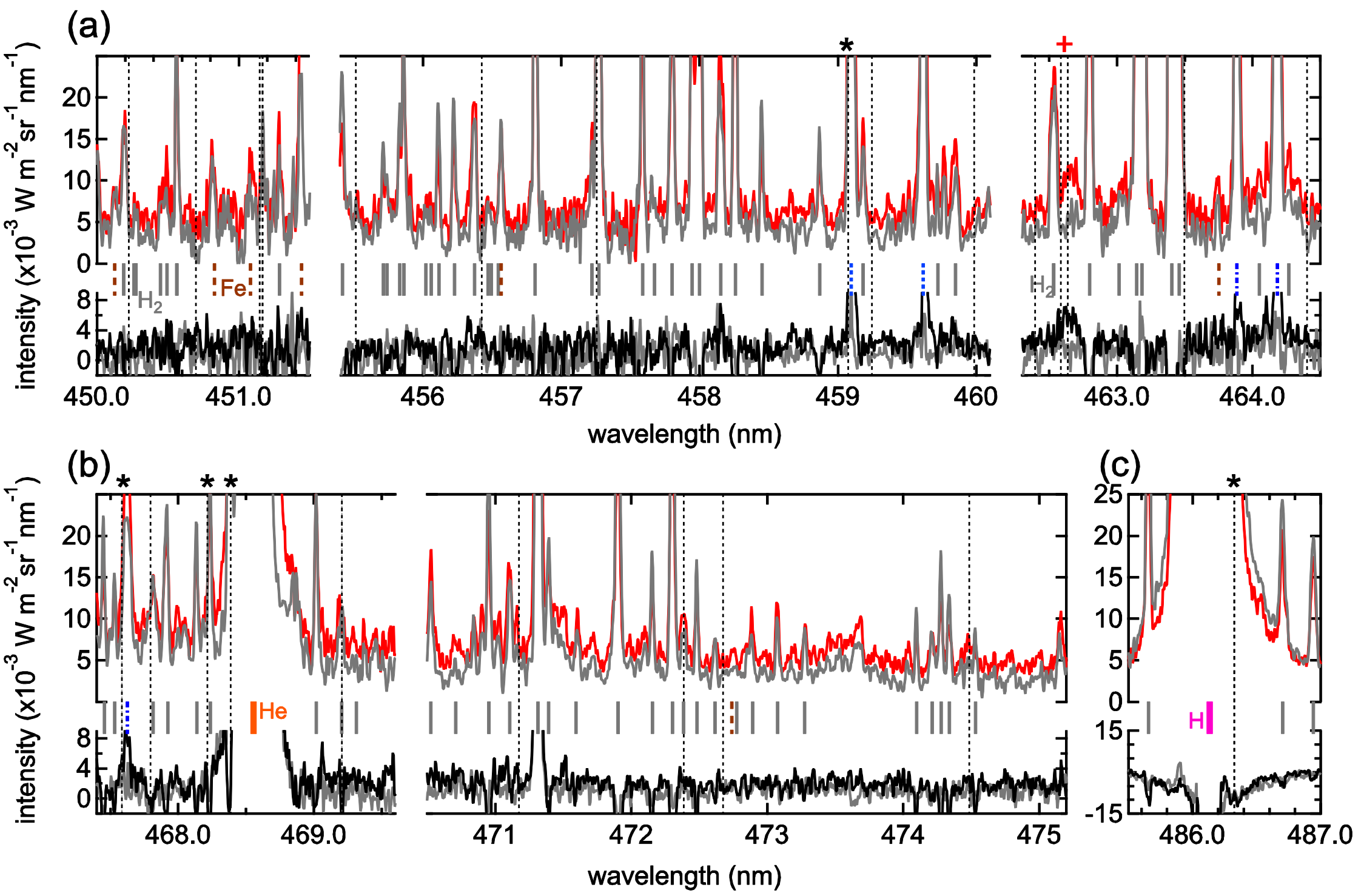

Fig. 5 (Upper panel) Emission spectra observed from LOS2 before (gray curves) and after (red curves) pellet injection. (Lower panel) Differential spectra for LOS1 (gray curves) and LOS2 (black curves). The central wavelengths of the emission lines observed in the EBIT are indicated by the vertical dotted lines. The emission lines also detected in this work are indicated by the vertical dotted lines with a marker (+) above. The vertical dotted lines with a marker (*) above indicate the emission lines that are not detected in this work because of the overlap by other intense emission lines. The other vertical dotted lines show the emission lines that are not detected in this work. The center wavelengths of the emission lines discovered in this work are given by the vertical blue chain lines.

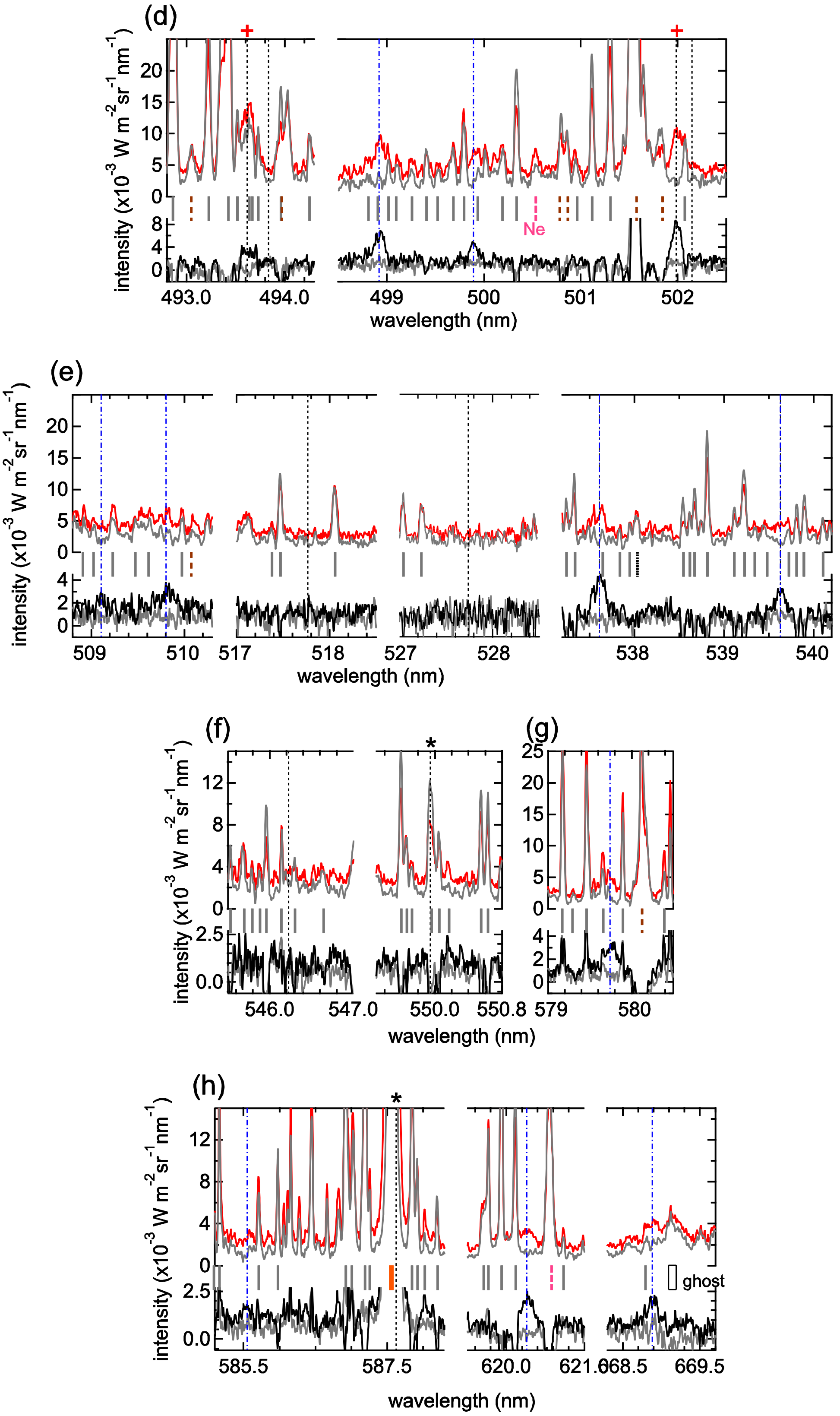

Fig. 5 (continued)

## 4. Conclusions and Discussion

Visible emission from LHD plasma, in which tungsten ions accumulated, was observed in the 450–715 nm wavelength range, with <0.05 nm wavelength resolution, simultaneously by our custom-built échelle spectrometer. 12 unknown emission lines were discovered. Because these 12 emission lines were observed (1) only after the tungsten pellet are injected into the plasma, (2) only from the core region of the plasma where the electron temperature is 1 keV, (3) with significant broadenings consistent with Doppler broadening from 1 keV tungsten ions, and (4) the wavelengths of some of these emission lines are close to those calculated by the Flexible Atomic Code [13,23], they can be attributed to highly charged tungsten ions. The precise determination of the emission location by the Abel inversion and $T_e$ value there will help the further identification of these emission lines.

Currently, it is unclear why some emission lines were detected in the EBIT experiment but not in LHD plasma and vice versa, as shown in table 1. A possible candidate is the difference of the plasmas; electrons in the EBIT experiment are mono-energetic, whereas both electron and proton energy distributions are present in LHD. Electron density in LHD plasma is also much higher than in the EBIT experiment.

**Acknowledgements**

This work was supported by the National Institute for Fusion Science (NIFS13KLPH021 and NIFS12KOAH028).

## Appendix

*Development of an Échelle spectrometer*

In this appendix, we describe the details of our custom-built échelle spectrometer that was used in this work.

A schematic illustration of the échelle spectrometer and our definition of the *xyz*-axes are shown in Fig.6. The ends of the two fibers are aligned along an entrance slit (SL1, aligned parallel to the sheet, 25 μm width). A close-up illustration around SL1 and the optical fibers are shown in Fig.6 (b). Emissions introduced through SL1 are collimated by a camera lens (L1, Nikon ED180mmF2.8D(IF), focal length: 180 mm; F number:2.8) and incident on an échelle grating (Richardson grating, ruling density: 52.67 grooves/mm; blaze angle: 63.5°; ruled area 46 × 92 $mm^2$). The collimated light beams are dispersed along the *z* direction and focused on another slit (SL2, aligned perpendicular to the sheet, 1 mm width). SL2 is adopted for the purpose of minimizing the stray light caused by the imperfection of the grating surface. The light passing through SL2 is collimated again by a camera lens (L2, ED180mmF2.8D(IF)) and incident on a cross disperser (Fig.6 (c)). The cross disperser consists of a custom-made prism (substrate: S-TIH1; apex angle: 30°; length of the hypotenuse: 58 mm; height: 60 mm) and a transmission grating (Richardson grating, ruling density: 100 grooves/mm; ruled area: 52 × 52 $mm^2$). The collimated light beams are dispersed by the cross disperser along the *y* direction. The light beams are focused on a complementary metal-oxide semiconductor (CMOS) image sensor of a digital camera (Hamamatsu Photonics, Orca Flash-4.0, pixel size: 6.5 × 6.5 $m^2$; 2048 × 2048 pixels). A schematic illustration of the dispersion by the échelle grating and the cross disperser is shown in Fig.7(a).

An example of the CMOS image and its expanded one obtained for an LHD discharge with an exposure time of 500 ms is shown in Fig.7 (b) and (c), respectively. As indicated by the red stripes in Fig.7 (c), the edge images of the two optical fibers are focused separately. The emission intensity of each *z* pixel is calculated by integrating over *y* pixels in the red stripe (software binning).

Conversion to wavelength from the *z* pixel is performed by using many visible emission lines of thorium and argon atoms in a hollow-cathode discharge lamp (Photoron, P858A). The instrumental widths of the system are also evaluated from the observed widths of the thorium and argon emission lines. The full widths at half maximum (FWHM) of the instrumental width as a function of the wavelength are shown in Fig.8. Since the achromatic aberration of the camera lenses are not perfectly corrected, the widths depend on the wavelength (e.g., larger instrumental widths for $\lambda \sim$ 550 nm, and $\lambda > 650$ nm). The sensitivity of each stripe is calibrated against a standard halogen lamp with an integration sphere (Labsphere, USS-600C).

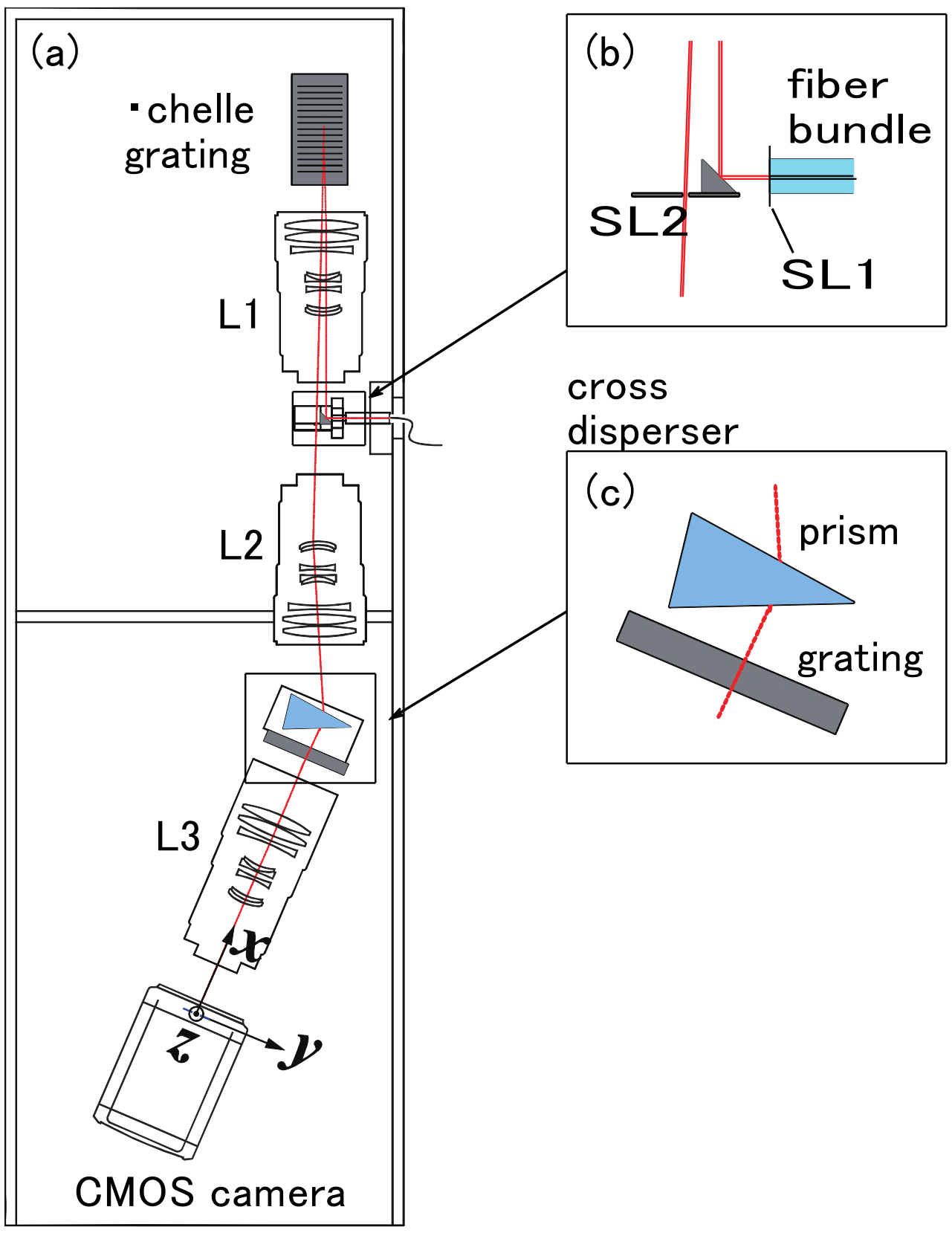

Fig.6 (a) A schematic illustration of the échelle spectrometer. Close-up views (b) near the entrance slit and (c) near the cross disperser. The échelle grating disperses the light along the *z* direction, whereas the cross disperser separates the orders of the diffracted light in the *y* direction.

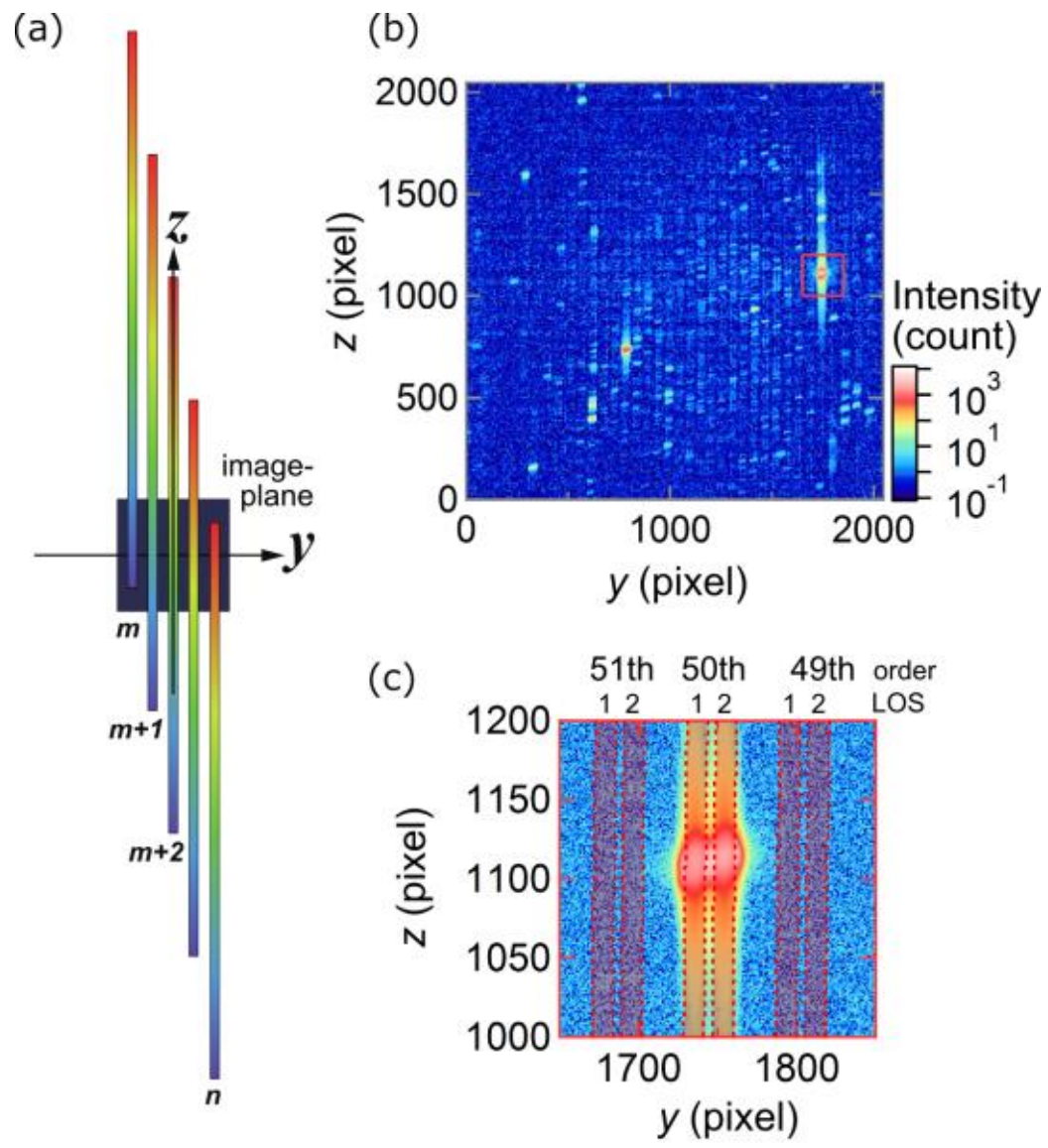

Fig.7 (a) A schematic illustration of the diffracted light by the échelle grating and the cross disperser focused on the image sensor. (b) An observed image of emissions from a hydrogen LHD plasma. (c) The expanded image of the region indicated by the red square in (b). The focused positions of the 49th, 50th, and 51st orders light introduced by the two optical fibers (LOS 1 and 2) are indicated by red stripes. The two bright spots located around the center of the figure are due to the Balmer-α emission ( $\lambda$ = 656.28 nm).

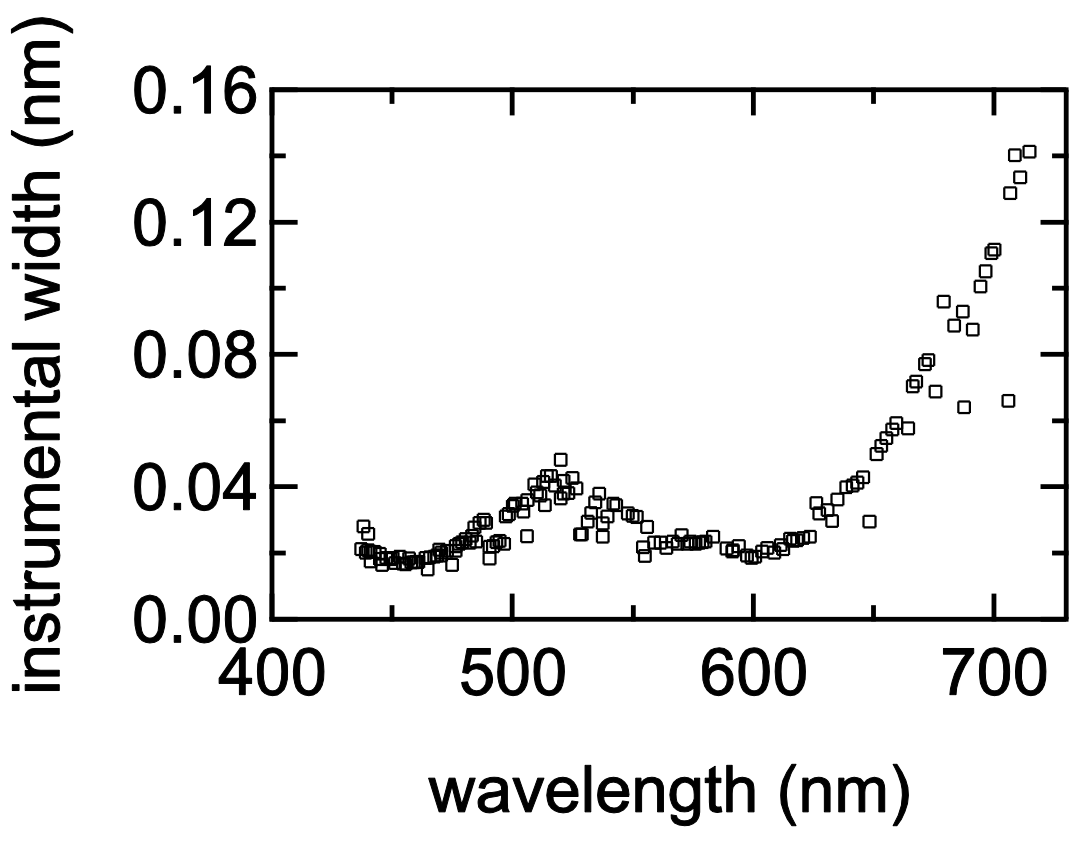

Fig.8 The wavelength dependence of the instrumental widths.